**ROGERIO BONATTI**
**ARTHUR GOLA DE PAULA**

# Development of email classifier in Brazilian Portuguese using feature selection for automatic response

**ROGERIO BONATTI**

**ARTHUR GOLA DE PAULA**

# Development of email classifier in Brazilian Portuguese using feature selection for automatic response

Monograph presented to the Department of Mechatronics Engineering and Mechanical Systems of Escola Politécnica da Universidade de São Paulo for obtention of the degree of Engineer

Advisor: Prof. Fabio G. Cozman

São Paulo

2015

**ROGERIO BONATTI**

**ARTHUR GOLA DE PAULA**

# Development of email classifier in Brazilian Portuguese using feature selection for automatic response

Monograph presented to the Department of Mechatronics Engineering and Mechanical Systems of Escola Politécnica da Universidade de São Paulo for obtention of the degree of Engineer

Concentration area: Mechatronics Engineering

Advisor: Prof. Fabio G. Cozman

BONATTI, R; PAULA, A.G.; **Development of Email Classifier in Brazilian Portuguese Using Feature Selection for Automatic Response**. São Paulo. 2015. 30 p. Graduação Escola Politécnica, Universidade de São Paulo, São Paulo, 2015.

## ERRATUM

| PAGE | LINE | WHERE YOU READ | YOU SHOULD READ |
|---|---|---|---|

## ACKNOWLEDGEMENTS

We would first like to thank our parents for all the support, love, education and encouragement they have given us, what allowed us to achieve what we have today.

We also thank *Fundação Estudar*, a Brazilian non-profit organization in the field of education, for providing all the email data utilized in the study.

We thank our advisor, Prof. Fabio G. Cozman for the guidance throughout our work, and for helping us achieve our full scientific potential.

We thank Carlos Henrique Pergurier and Cândido Leonelli for helping us find practical and impactful applications for our study and connecting us with scientific and business leaders.

We also acknowledge the help and interest of Victor Lamarca throughout the project.

## ABSTRACT

Automatic email categorization is an important application of text classification. We study the automatic reply of email business messages in Brazilian Portuguese. We present a novel corpus containing messages from a real application, and baseline categorization experiments using Naive Bayes and Support Vector Machines. We then discuss the effect of lemmatization and the role of part-of-speech tagging filtering on precision and recall. Support Vector Machines classification coupled with non-lemmatized selection of verbs, nouns and adjectives was the best approach, with 87.3% maximum accuracy. Straightforward lemmatization in Portuguese led to the lowest classification results in the group, with 85.3% and 81.7% precision in SVM and Naive Bayes respectively. Thus, while lemmatization reduced precision and recall, part-of-speech filtering improved overall results.

# RESUMO

Classificação automática de emails é um importante problema de classificação de textos. Nós estudamos e implementamos um classificador de emails no contexto empresarial em português brasileiro. Nós apresentamos um novo corpo de emails contendo estas mensagens, e um sistema de classificação para referência baseado nos classificadores Naive Bayes e Support Vector Machines. Nós também discutimos os efeitos de lematização e filtro por classe morfológica nos resultados de precisão e recall. O uso de Support Vector Machines associado à seleção morfológica de verbos, substantivos e adjetivos em texto não lematizado foi a melhor combinação, com 87.3% de acurácia. Apenas lematização do vocabulário em Português levou aos piores resultados encontrados no grupo, com 85.3% e 81.7% de precisão com Support Vector Machines e Naive Bayes respectivamente. Além disso, enquanto lematização reduziu precisão e recall de modo geral, o filtro de classe morfológica melhorou os resultados.

## DECLARAÇÃO DE ORIGINALIDADE

Este relatório é apresentado como requisito parcial para obtenção de grau de engenharia mecatrônica na Escola Politécnica da Universidade de São Paulo. É o produto do nosso próprio trabalho, exceto onde indicado no texto. O relatório pode ser livremente copiado e distribuído desde que a fonte seja citada.

## LIST OF ABBREVIATIONS

| | |
|---|---|
| **F1** | Harmonic mean between precision and recall |
| **IDFT** | Inverse Document Frequency Transform |
| **kNN** | k-Nearest-Neighbors |
| **NB** | Naive Bayes |
| **NLP** | Natural Language Processing |
| **POS** | Part of Speech |
| **PR** | Precision |
| **REC** | Recall |
| **SVM** | Support Vector Machines |

**CONTENTS**

# 1. INTRODUCTION

Electronic mail is an ubiquitous mode of communication in personal and work life [1], [2]. Email is quickly received, and it can be sent asynchronously at low cost. On the other hand, providing personalized and appropriate answers to questions sent by email is not an easy task, particularly as the number of messages scales up [3]. Messages are written in natural language and may contain several questions concatenated in a single sentence, or even implicit questions, perhaps containing ambiguous terms. Automatic replies are particularly useful in enterprises and institutions that receive hundreds or thousands of emails per day regarding specific categories such as products or divisions. Incoming messages can be separated by subject prior to reaching an employee, saving analysis time, expediting the answer and potentially increasing the answer's accuracy.

Several techniques have been developed [3]–[5] to automatically identify questions and intents in an email input, so as to either automatically answer questions or forward the message to an expert. These techniques are often based on Natural Language Processing (NLP) text understanding tools.

A common approach to text understanding is to classify incoming text into categories that are previously specified over the domain of interest. While the first applications of machine learning to email filtering appeared in the context of spam filtering, classification methods can be applied to message filtering into user-defined folders, automatic forwarding to other addresses in companies with subject sectorization, and to automatic replies [6]. A major difference between spam detection and classification of email messages for automatic answering is the number of categories: while the former application has two categories, the latter application usually deals with dozens or even hundreds of potential classes depending on the complexity of the organization. Indeed this is the sort of challenge we discuss in this thesis.

One can use machine learning algorithms to automatically learn classification rules based on training data that were previously classified by hand, in a supervised learning process [6]. Usually the accuracy of resulting classifiers is dependent upon the quantity of training data available [7], [8]. Often one combines labeled and unlabeled data [9], [10]; this thesis we focus on supervised learning only, leaving the use of unlabeled data to future work.

There are many techniques that can be applied to email and text classification, such as k-Nearest-Neighbors (kNN) [2], Decision Trees, Support Vector Machines (SVM) [11], [12] and Naïve Bayes (NB) classifiers. The latter two will be further detailed in Section 2. State-of-the-art algorithms vary depending on the type of classification being performed, that could be binary or between multiple categories, text length and the types of features to be taken into account in the statistical method [13], [14].

In this project we examine the problem of automatic email classification in multiple categories for messages written in Brazilian Portuguese. Even though text understanding and binary email classification have been explored in literature, very little work has been published on multi-categorical email classification for Portuguese. An exception is the work of Lima [15], who describes work on binary email classification in Portuguese by exploring differences among multiple algorithms, but provides few details over the types of tests, dataset characteristics and results obtained in the case of classification over multiple categories. We report our preliminary efforts in dealing with automatic email answering in Portuguese.

We have been driven to this problem by observing the business automation needs concerning customer service interaction in companies and institutions that receive hundreds of messages per day, in most part processed manually and inefficiently considering current NLP technology. Considering that 50% of today’s calls to call-centers fail to fulfill their objectives [16], an automated email classifier / response system could reduce the number of messages to be addressed by a human operator, thus reducing operational costs and response time. In addition, an automated classification system eliminates emotional and biological factors that can diminish the precision of manual classification, such as illnesses, mood changes and tiredness.

## 1.1. PURPOSE

Our goals are:

a) To build a corpus, containing business client interaction messages in Brazilian Portuguese, large enough for training / testing of statistical classification methods;
b) To evaluate the automatic email classification in Brazilian Portuguese of this dataset with the Naive Bayes and Support Vector Machines, as a baseline for future exploration;
c) To evaluate the impact, on recall and precision, of pre-processing incoming messages with a lemmatizer;
d) To evaluate the impact, on recall and precision, of a part-of-speech tagger feature selector.

## 1.2. THESIS' ORGANIZATION

This thesis is organized as follows. A brief literature review on email classification in folders and spam detection is presented in Section 2. The corpus collection procedure is explained in Section 3, and a description of the corpus processing for the experiments is given in Section 4. Results are presented in Section 5. Finally, Section 6 discusses the results obtained and proposes possible future work.  Section 7 concludes the thesis.

# 2. RELATED WORK AND BACKGROUND

## 2.1. NATURAL LANGUAGE PROCESSING AND TEXT CLASSIFICATION

The most accurate algorithms for text classification today are Support Vector Machines (SVM), Naive Bayes (NB) and k-Nearest-Neighbors (kNN), including hybrid approaches that can achieve greater precision than these methods separately [14]. SVM is one of the top performers for longer texts, but may present problems with shorter snippets [13]. SVM is usually implemented with linearly separable text, *i.e.*, binary classification like spam vs. ham or sentiment analysis like positive vs. negative. Multi-categorical uses are also possible, and are usually solved by using a sequence of binary classifications of the type one-versus-rest [17]. The NB method relies on the frequency of a word in the text. Both techniques will be further explained later in this section.

Our problem presents a context of high dimensional feature space and multi-categorical classification. This thesis focuses on the SVM and NB methods only, due to their robustness [18] to deal with these constraints.

Text classification algorithms typically take into account three types of features extracted from emails: unstructured text, categorical text and numerical data [19]. Unstructured text consists of the subject line and corpus, usually grouped in a "bag of words", while categorical data is well defined, and can be found in the sender and recipient domains for instance. Numerical data is related to message size and number of recipients. Experiments in the literature have concluded so far that numerical data are not very useful for email classification [12]. Additionally, feature selection filters may be applied to reduce noise in document classification and also to reduce the vocabulary used in computations.

Classifiers may use features based on word complexity, part-of-speech (POS) tags and presence of alphanumeric characters to enhance classification [20]. Lemmatizers can also be used in text categorization to treat different variation of the same root words as one, bringing verbs to the infinitive form for example.

## 2.2. NAIVE BAYES CLASSIFIER

The Naive Bayes classifier is a probabilistic classifier often applied to text categorizations tasks [21], [22]. Suppose each email instance (the evidence) *d* is described by a conjunction of word attribute values $\langle t_1, t_2, \ldots, t_{nd} \rangle$, and that are m target classes $\langle c_1, c_2, \ldots, c_m \rangle$. The probability of a document *d* to belong to class *c* is given in eq.1 [6]:

$$P(c|d) = \alpha P(c) \prod_{k=1}^{nd} P(t_k|c) \quad (1)$$

Where $\alpha$ is a constant, $P(tk|c)$ is the conditional probability of feature $tk$ occurring in a document of class C. We interpret $P(tk|c)$ as a measure of how much feature $tk$ contributes that *c* is the correct class. $P(c)$ is the prior probability of a document occurring in class c. The best class for the document is computed as given in eq.2 as the class highest probability (*maximum a posteriori)*[6]*:*

$$c_{chosen} = \arg max\, \hat{P}(c|d) = \arg max\, \hat{P}(c) \prod_{k=1}^{nd} \hat{P}(t_k|c), \quad c = 1, \ldots, c_m \quad (2)$$

Where $\hat{P}(c|d)$, $\hat{P}(c)$ and $\hat{P}(tk|c)$ represent the values of the parameters extracted from the training corpus from their relative frequencies (following a multinomial distribution).
The Naive Bayes algorithm proves to be a very computationally efficient [6] and precise [23] method for classifying texts into categories, despite the overly simplistic approach of assuming complete independence between words in a sentence, what does not even take into account the order of words in a text.

A simple numerical example to facilitate the understanding of the Naive-Bayes classifier was extracted from Manning, Raghavan and Schutze [6] and reproduced below:

**Table 1: Data for parameter estimation example. Extracted from Manning, Raghavan and Schutze (6)**

| | docID | Words in Document | In c=China? |
|---|---|---|---|
| Training set | 1 | Chinese Beijing Chinese | Yes |
| | 2 | Chinese Chinese Shanghai | Yes |
| | 3 | Chinese Macao | Yes |
| | 4 | Tokyo Japan Chinese | No |
| Test set | 5 | Chinese Chinese Chinese Tokyo Japan | ? |

*To estimate the probability of document $d5$ belonging to the class "China", we first compute the parameters $\hat{P}(c) = 3/4$ and $\hat{P}(\bar{c}) = 1/4$, and then the conditional probabilities for each one of the words:*

$$\hat{P}(Chinese|c) = \frac{5+1}{8+6} = \frac{6}{14} = \frac{3}{7}$$

$$\hat{P}(Tokyo|c) = \hat{P}(Japan|c) = \frac{0+1}{8+6} = \frac{1}{14}$$

$$\hat{P}(Chinese|\bar{c}) = \frac{1+1}{3+6} = \frac{2}{9}$$

$$\hat{P}(Tokyo|\bar{c}) = \hat{P}(Japan|\bar{c}) = \frac{1+1}{3+6} = \frac{2}{9}$$

*And finally we calculate the probability of each possible class given document $d5$:*

$$\hat{P}(c|d5) \propto \frac{3}{4} * \left(\frac{3}{7}\right)^3 * \frac{1}{14} * \frac{1}{14} \approx 0.0003$$

$$\hat{P}(\bar{c}|d5) \propto \frac{1}{4} * \left(\frac{2}{9}\right)^3 * \frac{2}{9} * \frac{2}{9} \approx 0.0001$$

*Therefore we assign document $d5$ to class $c = China$.*

## 2.3. SUPPORT VECTOR MACHINES CLASSIFIER

Support vector is a classification method developed by Cortes and Vapnik [24] based on calculating a hyper plane on a high dimensional space that achieves the largest separation from the data points [18]. Such a problem can be qualitatively viewed in Figure 1, which was extracted from Joachims [18].

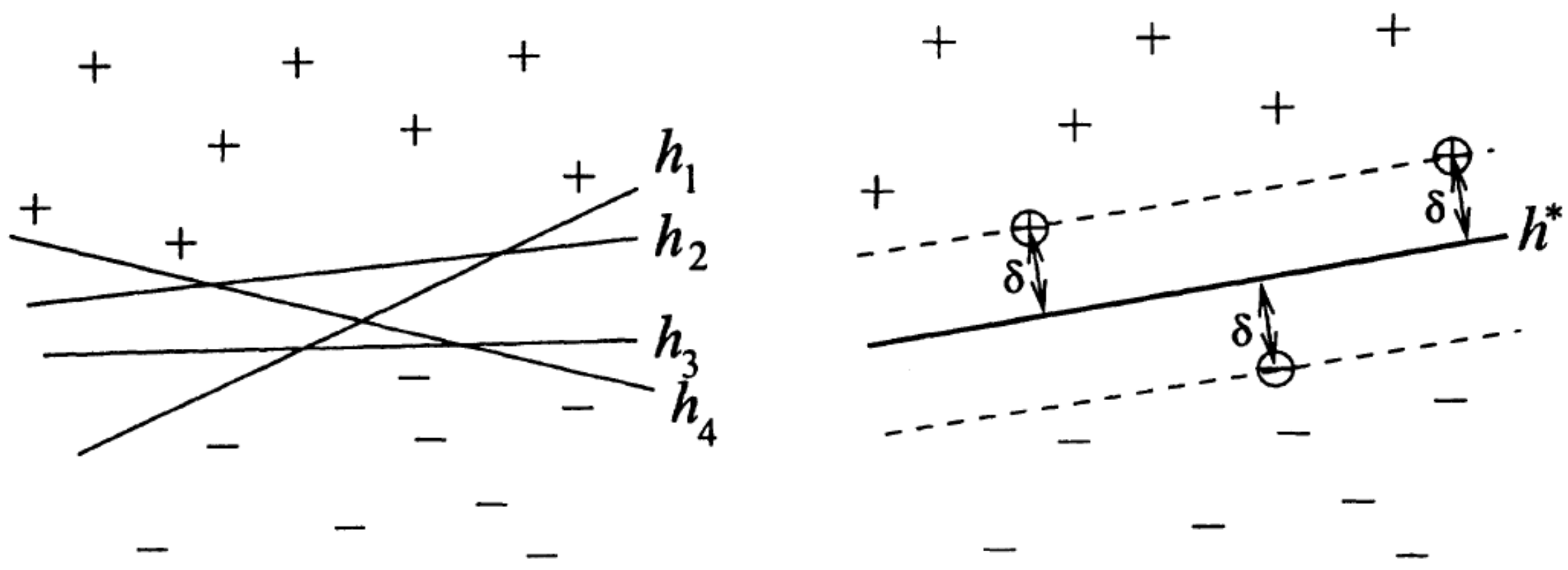

**Figure 1: A binary classification problem in two dimensions. Positive examples are marked by +, negative examples by -. Left: many hyperplanes separate the training examples without error. Right: support vector machines find the hyperplane *h**, which separates the positive and negative training examples with maximum margin $\delta$. The examples closest to the hyperplane are called support vectors (marked with circles) [18]. Extracted from Joachims [18].**

Given instance-label pairs $(x_i, y_i), i = 1, \ldots, l$ where $x_i \in \mathbb{R}^n$ and $y \in \{1, -1\}^l$, the optimization problem to be solved is [25]:

$$min_{w,b,\varepsilon} \frac{1}{2} w^T w + C \sum_{i=1}^{l} \varepsilon_i \quad (1)$$

subject to $y_i(w^T \phi(x_i) + b) \geq 1 - \varepsilon_i$, with $\varepsilon_i \geq 0$

Where $w$ is the vector normal to the hyperplane (not necessarily normalized), $b$ determines the offset from the hyperplane from the origin, $C > 0$ is the penalty parameter of the error term and $\varepsilon_i$ measures the degree of misclassification for $x_i$ [25].

In many applications, the relationship between points from different classes may be non-linear, therefore a non-linear mapping function ϕ should be used. As shown by Boser et al. [26], a kernel function $K(x_i, y_i)$ can be used to compute the scalar product between, as given by $K(x_i, y_i) = \phi(x_i)^T \phi(x_i)$. In practice, four basic kernels are commonly used by researchers [27], [25]:

- Linear $K(x_i, x_j) = x_i{}^T x_j$
- Polynomial $K(x_i, x_j) = (\gamma x_i{}^T x_j + r)^d,\ \gamma > 0$
- Radial Basis Function (RBF) $K(x_i, x_j) = \exp\left(-\gamma \left\|x_i - x_j\right\|^2\right),\ \gamma > 0$
- Sigmoid $K(x_i, x_j) = \tanh\,(\gamma x_i{}^T x_j + r)$

Where $\gamma, r$ and $d$ are kernel parameters, and can be optimized for better classifier performance. For the case of text classification it is empirically recommended to use simple linear kernels with the data, since the number of features is much higher than the number of instances [25].

## 2.4. EVALUATION OF CLASSIFIERS

To evaluate the performance of classifiers, in particular to compare classifiers with non-lemmatized training set versus lemmatized training sets, three parameters are calculated: precision, recall and accuracy. These measures and their qualitative meanings are given in eq. 3-5 [2]. These quantities are expressed in terms of True Positives (TP), True Negatives (TN), False Positives (FP) and False Negatives (FN) for each individual class. The previous four parameters can be better visualized in a contingency table (Table 2), which shows their relevance for the document retrieval system [6]:

**Table 2: Contingency table for document retrieval system. Source: from Manning, Raghavan and Schutze (6).**

| | Relevant | Non Relevant |
|---|---|---|
| Retrieved | True Positive (TP) | False Positive (FP) |
| Not Retrieved | False Negative (FN) | True Negative (TN) |

### 2.4.1. Precision (PR) measure

Precision is the fraction of positive predictions that are relevant in a given class or cluster of classes, given by eq. 3. It measures the certainty that we have that what we classified is correct, ignoring if some of the relevant documents were not retrieved.

$$\frac{TP}{TP + FP} \qquad (3)$$

### 2.4.2. Recall (REC) measure

Recall is the fraction of positive-labeled instances that are retrieved in a given class or cluster of classes, given by eq. 4. It evaluates the ability of the classifier to retrieve all relevant documents, accepting retrieval of wrong documents.

$$\frac{TP}{TP + FN} \qquad (4)$$

### 2.4.3. F1 measure

F1 is the harmonic mean of precision and recall, given by eq. 5. It is commonly used to assess classifiers more generally with one single measurement taking into account precision and recall at the same time.

$$\frac{2\ PR\ REC}{PR + REC} \qquad (5)$$

## 2.5. LEMMATIZER

To avoid treating different inflections of the same word as distinct attributes in the statistical counting, a lemmatizer can be used to bring all words to their lemmas in the corpus. An example of application of the lemmatizer is presented in Table 3.

**Table 3: Example of application of the lemmatizer used to filter text.**

| Language | Original text | Lemmatized |
|---|---|---|
| PT | Onde é a inscrição no programa de bolsas? | Onde ser o inscrever em programa de bolsa? |
| EN | Where is the enrollment in the scholarship programs? | Where be the enrollment in the scholarship program? |

## 2.6. PART-OF-SPEECH FILTERING

POS Tagger filter can be applied to the studied corpora to remove classes of words considered irrelevant or noise to text classification (such as verbs, nouns adjectives, etc.). As discussed in literature [28], it may be advantageous to use POS tags in text classifiers because:

- Information retrieval with POS tags improves the quality of the analysis in many cases [29].
- It is a computationally inexpensive method to increase relevance in the training set.

An example of a POS filtered phrase is given in Table 4.

**Table 4: Example of application of the Part-of-Speech filter to text, leaving only verbs and nouns.**

| Language | Original text | POS filtered text (only verbs and nouns left) |
|---|---|---|
| PT | Onde é a inscrição no programa de bolsas? | é inscrição programa bolsas |
| EN | Where is the enrollment in the scholarship programs? | is enrollment scholarship programs |

## 2.7. RELATED WORK

### 2.7.1. Text classification

There are many examples of email classification using machine learning algorithms in literature. The most common applications of email sorting are in the field of spam filtering.

Crawford et al. [30] compiled in 2001 some of the first classification systems such as the work of Androutsopoulos [23] which achieved 95% precision and 78% recall with a Naive Bayes classifier for spam filtering, and the work of Provost [31] that achieved 95% accuracy also using NB for a binary classification of spam. Modern spam filters can achieve more than 95% recall and precision together [32], being extremely efficient and precise in binary classification.

Email and short text categorization is also applied to multi-topic selection, such as separating emails in folders or classifying emails per gender. Klimt and Yang [12] presented an email classification system in folders using the Enron Dataset with an F1 score near 0.7 using a SVM classifier. Chen et al. [33] worked with microblog messages

such as *Twitter*, classifying them into 6 categories like Sports, Business, etc., achieving both precision and recall close to 80%. Microblog messages are similar to emails in the sense that they use colloquial language and present relatively short sentences.

In the Portuguese language, binary classification algorithms practically achieve state-of-the-art levels. Silva [34] and Moreira [35] presented spam classifiers with true positive rates above 99%. Also in the field of short document analysis, Santos [36] classified online product reviews as positive or negative with 78% precision and 81% recall.

Lima [15] produced significant results on the topic of business email classification in Portuguese, comparing the performance of different classifiers on a set of emails labeled in folders. Lima presents F1 scores around 90% for binary classification in folders using kNN, and achieving 76% precision and 81% F1 for multi-topic classification with SVM. However, Lima provides few details on the reasons why SVM outperformed other classifiers in terms of the specific dataset's characteristics, which is not publicly available.

### 2.7.2. Feature selection

On the topic of feature selection for text classification, several papers are worth mentioning.

In the micro blog context, Kouloumpis, Winson and Moore [37] classified Twitter messages into positive or negative using multiple linguistic features such as separating words in *n-grams*, lexicon polarity and part-of-speech tags in different combinations. Results showed that using POS tags as a word feature decreased classification accuracy, going from about 65% F1 in the best case to approximately 55% F1 when POS tags are applied. Work by Batool et al. [38] took a different approach of the use of filters: keywords were extracted from the text, and the best results were obtained with leaving only verbs and entities like hash tags in the text.

Pang, Lee and Vaithyanatha [39] tried a different approach in movie review sentiment analysis, by comparing the performance of NB and SVM classifiers with datasets containing all parts-of-speech versus solely using adjectives for classification. Their results showed that despite the apparent expectation that adjectives contain most of the information relative to the positivity or negativity of a movie review, the vocabulary limitation actually decreased classification performance from 82% to 77% in accuracy.

# 3. CORPUS COLLECTION

In this section we explain how we built the corpus we used in all experiments.

## 3.1. PARTNERSHIP WITH NOT-PROFIT ORGANIZATION TO OBTAIN DATA

Even though email communication is important in many interactions between customers and companies, real-life data is of limited availability. To our knowledge, no public enterprise email corpus with multiple labeled categories is now available in Brazilian Portuguese, so it was necessary to partner with a company to run our experiments. Indeed, this work started during conversations with an organization that offers online support to customers and that was interested in automatic email answering. We focused in negotiating a partnership with a company that possesses a Business to Consumer (B2C) relationship, due to the fact that such a relationship would allow us to gather a larger quantity of data for our database of previously asked questions in comparison with a Business to Business (B2B) relationship.

We formed a partnership with Fundação Estudar, a non-profit organization in the field of education, that offers services such as student funding, prep courses and entrepreneurship workshops. They receive an average of 200 emails every day and agreed to share their database with us. The interactions are mainly with customers asking questions about their services or requesting support.

We gathered a raw corpus containing 35,218 emails, with all emails written in Brazilian Portuguese. The raw corpus corresponds to all the email messaging interactions that *Fundação Estudar* had in six months, both incoming and outgoing. We chose not to collect data over a longer period, because there were significant changes in the institution's activities prior to this period, which might affect the classification negatively due to changes in message categories.

## 3.2. EMAIL IS STRUCTURED IN TICKETS

Messages in our corpus, as commonly seen in customer relation services, are structured in *tickets*. A *ticket* corresponds to two or more email exchanges over the same topic. Typically, a *ticket* starts with a first email from a costumer asking a question, requesting technical support or sometimes giving information to the institution. Customer relations staff then reply the first email. The reply is stored in the same *ticket* as second email. Nearly 75% percent of tickets end in two interactions and the rest store three or more. In cases where the same customer contacts the institution again in another email, an additional *ticket* is created to store the new conversation. We assembled 15,297 *tickets* in total.

Figure 2 shows an example of a typical ticket:

- First email is from a customer asking for information.
- Second email is the reply from the customer relation team.
- Third email is a thank you email from the customer that closes the ticket.

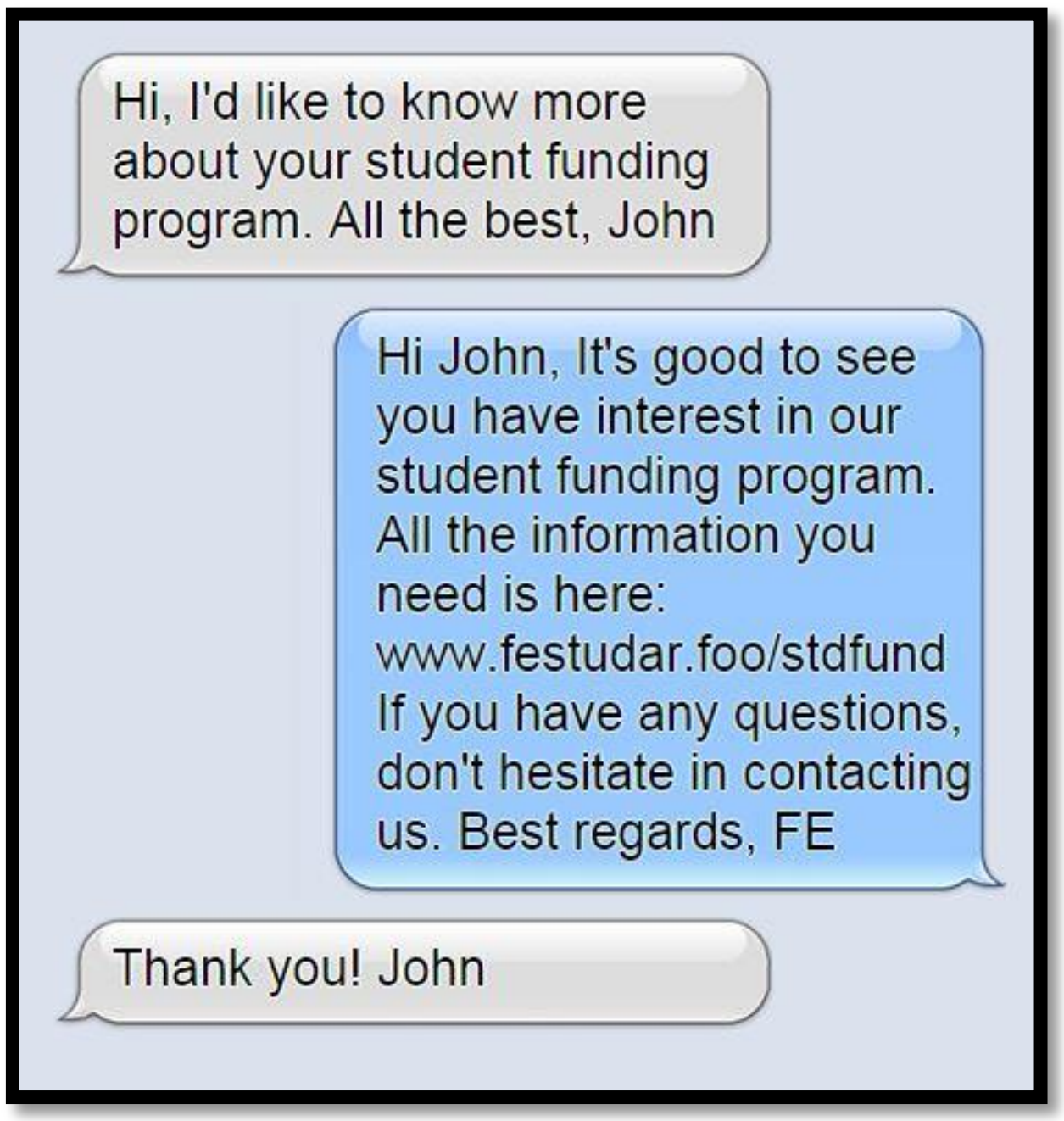

**Figure 2: Example of a ticket.**

## 3.3. MACROS AS CLASSES FOR MACHINE LEARNING ALGORITHM

*Fundação Estudar*'s customer relations staff use pre-written messages in their daily work of replying customers' emails, since most messages can be fit in predetermined categories and answered with minor changes. The pre-written messages, called *macros,* represent 120 types of frequently received emails on the products or products' subcategories from *Fundação*.

Our classifier uses a subset of the *macros* as *classes*. Correctly classifying an incoming email into one of the classes in the subset is necessary to reply to it automatically.
Table 5 shows an example of an incoming email, the standard macro to which the email belongs, and the actual answer employed by *Fundação Estudar*, which is a slightly modified version of the macro. The changes made by the staff are in bold.

**Table 5: Example of macro and how it is used, based on a real case.**

| First email | Macro used to reply the first email | Actual reply (adapted from macro) |
|---|---|---|
| Hi,<br><br>I want to study in the UK and I don't know how. Can you help me?<br><br>Regards,<br><br>Anna | Hello {XXX},<br><br>How are you?<br><br>In our webpage we have all the tips and a step-by-step guide of what to do. I'm sending you the main links, but do browse around the website to learn all you can about studying abroad.<br>=)<br><br>Guide: http://bit.ly/1BXKg0x<br>Selection: http://bit.ly/1nVnthr<br>Tests: http://bit.ly/1D2jUha<br>Recommendation letter : http://bit.ly/1DBypqp<br><br>Good luck!!! =) | Hello **Anna**,<br><br>How are you?<br><br>In our webpage we have all the tips and a step-by-step guide of what to do. I'm sending you the main links, but do browse around the website to learn all you can about studying abroad.<br>=)<br><br>Guide: *http://bit.ly/1BXKg0x*<br>Selection: *http://bit.ly/1nVnthr*<br>Tests: *http://bit.ly/1D2jUha*<br>Recommendation letter : *http://bit.ly/1DBypqp*<br><br>**A little bit more about the UK you can find here:** ***http://bit.ly/1Dnaklsl***<br><br>Good luck!!! =) |

## 3.4. EMAIL LABELING IN CATEGORIES

In our analysis, we considered tickets containing two or three email messages: one question email, one response email and one optional thank-you email. This decision was taken after studying sample of 20% of the tickets with more than three emails messages: we noticed that, in 45% of the times, the customer's third email was a secondary question, mostly because his/her first inquiry was not successfully replied. In the other 25% of the cases where tickets contained more than 3 messages, the ticket's first email was a personal question that was not clearly classifiable within the pre-determined categories, i.e., it could not be answered by a pre-written reply, and resulted in more than 1 interaction with Fundação Estudar. Therefore, to avoid using questions that were not correctly answered in our labeled data, meaning that we would have incorrectly classified inquiry emails in our dataset for machine learning algorithms, we opted to remove tickets with more than three emails from our study.

After the first triage, 11,410 *tickets* out of the original 15,297 remained. The next step of preparation of the corpus was the creating of our labeled data, obtained by labeling the remaining *tickets* within the *classes*, i.e., determining which *macro* could reply each of the emails. This was done by comparing the institution's answers with the pre-defined responses.

Considering that the staff makes small changes to the macros before using them, we defined the *core* of each *macro,* that is, the most important part of it that is not changed and that is not present in any other *macro*. We then tested each reply email, defined as the second email of the ticket, to find exact occurrences of the *cores* in them. The outcome of this process was a list of 2081 *tickets* in which the reply email corresponds to a *macro*, and therefore, the first email can be successfully replied by that *macro.*

These labels were used as labeled examples for the machine learning classification algorithms. Table 6 presents an example of the process used to extract the core of the same macro showed in Table 5.

**Table 6: Example of a macro, how it is used and its core for matching.**

| Original macro | Actual reply | Core defined for that macro |
|---|---|---|
| **Hello {XXX},**<br><br>**How are you?**<br><br>**In our webpage we have all the tips and a step-by-step guide of what to do. I'm sending you the main links, but do browse around the website to learn all you can about studying abroad.**<br>**=)**<br><br>**Guide: http://bit.ly/1BXKg0x**<br>**Selection: http://bit.ly/1nVnthr**<br>**Tests: http://bit.ly/1D2jUha**<br>**Recommendation letter : http://bit.ly/1DBypqp**<br><br>**Good luck!!! =)** | Hello **Anna**,<br><br>How are you?<br><br>In our webpage we have all the tips and a step-by-step guide of what to do. I'm sending you the main links, but do browse around the website to learn all you can about studying abroad.<br>=)<br><br>Guide: *http://bit.ly/1BXKg0x*<br>Selection: *http://bit.ly/1nVnthr*<br>Tests: *http://bit.ly/1D2jUha*<br>Recommendation letter : *http://bit.ly/1DBypqp*<br><br>**A little bit more about the UK you can find here: *http://bit.ly/1Dnaklsl***<br><br>Good luck!!! =) | In our webpage we have all the tips and a step-by-step guide of what to do. I'm sending you the main links, but do browse around the website to learn all you can about studying abroad. |

On average, there were 28 emails per class after matching the macros to the actual replies. The distribution of emails per class is depicted in Figure 3.

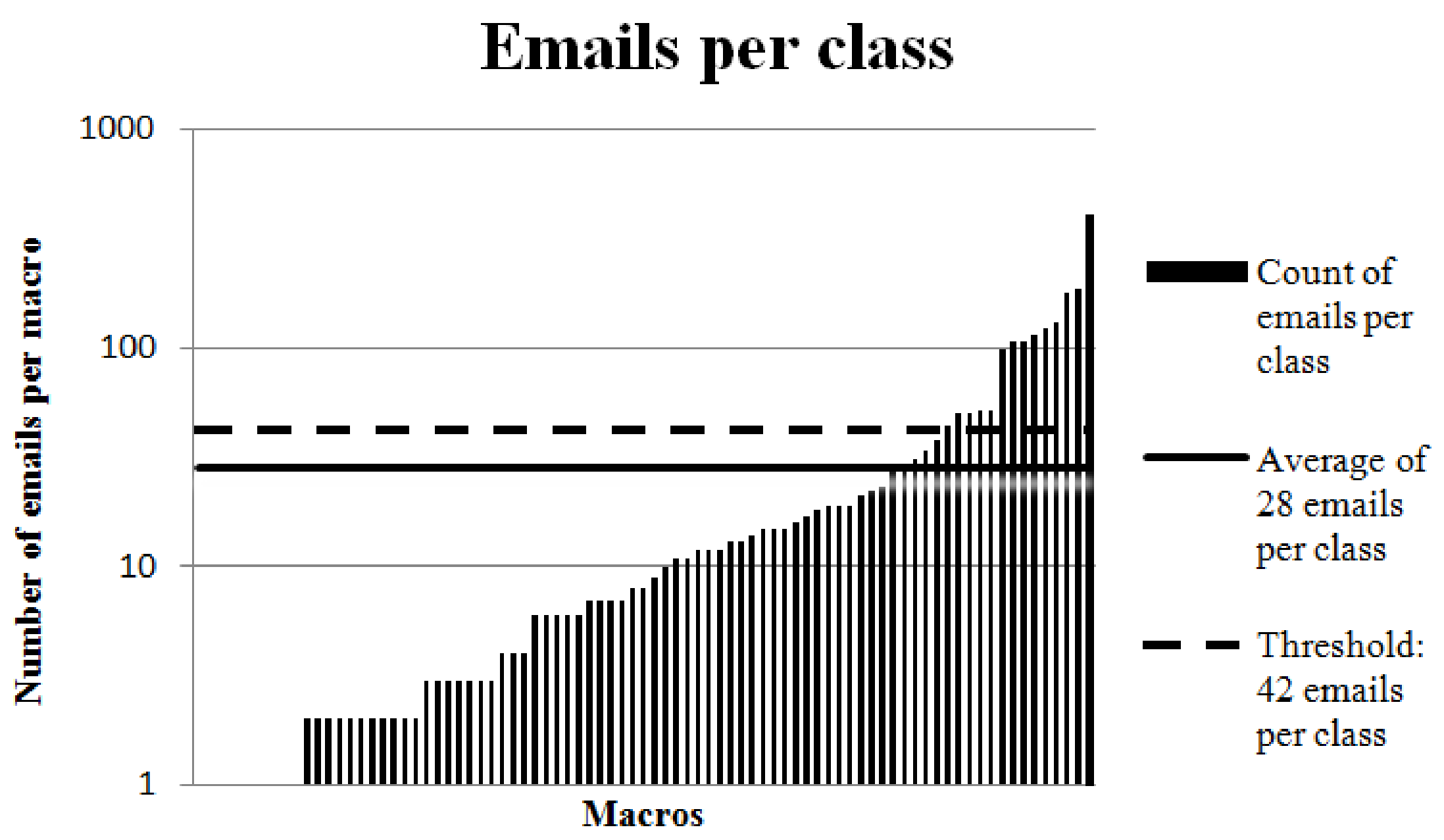

**Figure 3: Distribution of emails per class.**

We separated eight of the top 12 categories in number of emails for analysis, obtaining a total of 42 emails per class. We discarded four of these categories because they were generic answers that could refer to a variety of situations in the context of Fundação Estudar. Seven of the eight chosen categories had a number of emails larger than 42, but to balance the classes' vocabulary range and improve classification performance, we selected as much emails as the eighth class. Table 7 shows the names of classes for the classifier.

**Table 7: Classes for the machine learning algorithms.**

| Product category at Fundação Estudar | Message's subject |
|---|---|
| **In Practice / Immersion** | Received a participation confirmation in the program |
| **In Practice / Immersion** | Won't be able to participate in the program |
| **General** | Scholarship program info |
| **Error** | Forgot password - what to do? |
| **General** | Checking eligibility criteria for the program |
| **General** | Requesting feedback after application |
| **Error** | Solve enrolment errors in website |
| **Study Abroad Info** | How to study abroad: step-by-step |

## 3.5. TEXT AND EMAIL CORPORA IN LITERATURE

Other Portuguese language databases of manually annotated categories could be found, such as *Linguateca* [40] and *Floresta Sintática* [41], but they do not contain email messages. The work of Lima [15] contained an email corpus extracted from a private company in Portuguese, but it was not publicly available. In the English language there are several public corpora of labeled text belonging to more than 2 categories, such as the Reuters-21578 [30] corpus for news classification and the Enron [12] corpus for email classification, but we chose to study the classification of emails written in Portuguese, therefore we had to create our own corpus.

The corpus developed from Fundação Estudar's datalog is now in their possession, organized in text files by category.

# 4. CORPUS PROCESSING

In this section we explain how we processed our email corpus to prepare the datasets used in the experiments, and the techniques applied to classify messages.

## 4.1. TEXT FILTERING WITH LEMMATIZER AND PARTS OF SPEECH

We used different techniques to process the training corpus with the objective of assessing the impact on recall and precision of removing certain parts-of-speech and of lemmatizing the text of the messages. The first dataset separation was using a Brazilian Portuguese lemmatizer [42] to bring verbs to infinitive form and nouns and adjectives to the masculine and singular form. After this stage, the two corpora created, raw and lemmatized, were split into 16 groups by removing certain parts-of-speech and retaining others. The parts-of-speech were selected with a POS-Tagger for Brazilian Portuguese [43]. The filter configurations are shown in
Table 8.

**Table 8: Datasets created from raw corpus for the classification experiments.**

| Datasets | |
|---|---|
| **Lemmatizer applied** | **POS-Tagger filter** |
| No | No |
| No | Verbs and nouns without participle |
| No | Verbs and nouns only |
| No | Verbs nouns and adjectives |
| No | Verbs, nouns, adjectives and adverbs |
| No | Verbs, nouns, and relative pronouns |
| No | Verbs, nouns and conjunctions |
| No | Verbs, nouns and adverbs |
| Yes | No |
| Yes | Verbs and nouns without participle |
| Yes | Verbs and nouns only |
| Yes | Verbs nouns and adjectives |
| Yes | Verbs, nouns, adjectives and adverbs |
| Yes | Verbs, nouns, and relative pronouns |
| Yes | Verbs, nouns and conjunctions |
| Yes | Verbs, nouns and adverbs |

## 4.2. NAIVE BAYES AND SVM CLASSIFIERS

In our experiments, we used two different classifiers: Naive Bayes and Support Vector Machines. Distinct configurations for each of these algorithms were chosen taken into account the characteristics of our dataset.

For NB we opted for the multinomial configuration with Inverse Document Frequency (IDF) weighing for the vocabulary. These settings were chosen after literature review [6], [44] and preliminary tests with our dataset that showed its performance superior in relation to other options.

For the SVM classifier, we used a linear kernel due to the high dimensionality of our experiment with text classification. The linear kernel's superior performance for text is shown by Joachims [7] and Hsu and Chang [25]. Preliminary experiments showed that using IDF weighing diminished performance with SVM, therefore IDF was not used in the main experiment.

## 5. EVALUATION OF LEMMATIZATION AND PART-OF-SPEECH FILTERING EFFECT ON PERFORMANCE

Table 9 presents the effect of the POS-Tagger filter and of the lemmatizer in precision, recall and F1 measurements with our different training and test data. Comparing both classifiers among all filters, the highest precision achieved was 87.5%, recall 87.2% and F1 87.3%, for the training set containing verbs, nouns, adjectives and adverbs with unlemmatized emails and using linear-kernel SVM without IDF weighing. The results show that the lemmatizer reduces performance of the classifier, whereas the POS-Tagger improves it.

**Table 9: Effect of lemmatization and POS-Tagger filtering on precision (PR), recall (REC) and F1**

| Datasets | | Naive Bayes | | | SVM | | |
|---|---|---|---|---|---|---|---|
| **Lemmatizer** | **POS-Tagger** | **Precision** | **Recall** | **F1** | **Precision** | **Recall** | **F1** |
| No | No | 84.1% | 83.3% | 83.0% | 86.9% | 86.0% | 86.3% |
| No | Verbs and nouns without participle | 84.9% | 83.9% | 83.7% | 87.1% | 86.6% | 86.8% |
| No | Verbs and nouns only | 85.4% | 84.5% | 84.4% | 86.1% | 85.7% | 85.9% |
| **No** | **Verbs, nouns and adjectives** | **84.7%** | **84.2%** | **83.9%** | **87.5%** | **87.2%** | **87.3%** |
| No | Verbs, nouns, adjectives and adverbs | 84.3% | 83.3% | 83.1% | 87.1% | 86.6% | 86.8% |
| No | Verbs, nouns, and relative pronouns | 84.4% | 83.6% | 83.3% | 86.3% | 85.7% | 85.9% |
| No | Verbs, nouns and conjunctions | 85.1% | 84.2% | 83.9% | 86.7% | 86.3% | 86.4% |
| No | Verbs, nouns and adverbs | 83.0% | 82.1% | 82.1% | 85.1% | 84.5% | 84.6% |
| Yes | No | 83.4% | 82.4% | 82.4% | 84.7% | 84.2% | 84.3% |
| Yes | Verbs and nouns without participle | 83.8% | 82.7% | 82.8% | 83.8% | 83.3% | 83.4% |
| Yes | Verbs and nouns only | 83.6% | 82.4% | 82.4% | 84.5% | 83.9% | 84.0% |
| Yes | Verbs nouns and adjectives | 84.3% | 83.0% | 83.1% | 86.0% | 85.4% | 85.5% |
| Yes | Verbs, nouns, adjectives and adverbs | 83.5% | 82.4% | 82.5% | 83.2% | 82.7% | 82.8% |
| Yes | Verbs, nouns, and relative pronouns | 83.7% | 82.7% | 82.8% | 84.9% | 84.5% | 84.5% |
| Yes | Verbs, nouns and conjunctions | 82.5% | 81.5% | 81.5% | 85.0% | 84.5% | 84.6% |
| Yes | Verbs, nouns and adverbs | 84.1% | 83.3% | 83.0% | 86.9% | 86.0% | 86.3% |

# 6. DISCUSSION

## 6.1. THE CHOICE OF THE CLASSIFIERS

In this project, we focused on the Naive Bayes and SVM algorithms for classification. A common application of these classifiers is separating “Spam and Ham” in email inboxes [44], but they have also attained high precision and recall in classification problems with more than two categories [12]. Naive Bayes has been noted to rival SVM classifiers, often considered the state-of-the-art approach for text classification [14]. SVMs have shown superior results for sentiment analysis and other types of binary classifications [13], [14], but similar results to NB when dealing with sparse data and multi-topic classification. Our experiments show that, for email classification in Brazilian Portuguese SVM is significantly superior to NB.

## 6.2. PRECISION AND RECALL ARE CONSISTENT WITH LITERATURE

The values of precision and recall obtained in our experiments are similar to what is seen in literature for Naive Bayes email classification, or even general text classification. Our classifier reached precision, recall and F1 of 87.3%, above the range of 70 to 80% recall presented by Androusopoulos et al. [23] in binary classification for spam and ham.

On multi-category classification, Dewdney [45] tested different algorithms for seven very distinct categories and obtained, approximately, recall of 76% and precision of 80%. Chen et al. [33], who classified micro blog text within ten categories, reached 87% for both precision and recall.

## 6.3. COLLOQUIAL SPEECH REDUCE PERFORMANCE

One characteristic of our corpus that reduces performance is the fact that email messages are often an informal mean of communication. For example, when compared to a collection of newspaper articles as Reuters-21578 [30] that has much more vocabulary per text, longer texts and more formal language use, our corpus presents greater challenges for classification as these characteristics have great effect on the machine learning algorithm. In informal language, the reduced variety of words that are used results in a higher chance of finding two emails that have the same words and belong to different classes.

## 6.4. THE USE OF A LEMMATIZER IS NOT BENEFICIAL

Taking into account the fact seen in our experiments that the lemmatizer reduces performance of the classifier, the use of this particular NLP tool is not justifiable for our corpus, and, therefore, that for our dataset, verb tense information is relevant for classification and should not be removed.

The explanation to this experimental result comes from the definition of lemmatizing words, which is reducing them to their lemmas, and, therefore, losing the information that the words' inflections carry, such as verb tenses. An analogy can be made with a three-dimensional castle of cards. Suppose the castle of cards is a word. Lemmatizing the word would be the same as taking a photograph of the castle from the top: from the photo, it is still clear you are looking at cards, but you no longer understand they form a castle. Lemmatizing the words is losing a dimension of it, just like in the castle of cards. In our case as well as in our analogy, the dimension we lose represents loss in explanatory power.

**Table 10: Example of variation of verb tenses on different email subjects.**

| Subject of the email | Email | Relevant verb tense |
|---|---|---|
| Problem in Internet connection during online test for student funding selection process | Hi,<br>My Internet connection dropped during the test for the student funding selection process.<br>Can I re-take the test?<br>Thank you,<br>Louis | Past |
| Next phases of selection process for student funding | Hello,<br>I'd like to know when the next phases of the selection process for student funding will take place.<br>All the best,<br>Julian | Future |

Table 10 reveals the importance of verb tenses in our context in two examples. The first is an email about a problem had during an online test and the other, asking about the next phases of a selection process for student funding. In the first case, the verbs are most likely to be in the past, whereas for the second case, verbs tend to be in the future.

## 6.5. PART-OF-SPEECH FILTERING IMPROVES CLASSIFICATION

The experiments we carried out showed significant increase in performance of the classifier for POS-filtered datasets, which suggests that, in our context, nouns and verbs are the most significant parts-of-speech for the classification. A possible explanation for the significance for classification may come from the retained POS having better defined patterns for each class, considering our dataset size. The parts-of-speech removed (prepositions, conjunctions, pronouns, etc.) would then, act as noise in the classification.

Removing certain POS is reducing the information carried by the models for classification as well as using the lemmatizer, but for the POS, the filtered parts did not add relevant information to the classification. This phenomenon is specific for our dataset in terms of both context and size. In the context of email classification for costumer relations, nouns and verbs appear to carry the most relevant information, which may not be true for text classification in other contexts. In sentiment analysis, for example, adjectives and adverbs are likely to have greater importance.

## 6.6. NEXT STEPS

The main future developments for this project are:

- Improving labelling precision by increasing the number of emails per class for the machine learning process, therefore gathering more labelled email data referring to each category from Fundação Estudar's records;
- Increasing the number of email categories in the classification scheme so as to have a more comprehensive email classifier applicable to real life situations with dozens of categories, therefore selecting more non-overlapping email categories in Fundação Estudar's responses;
- Creating a user interface to apply the email classification techniques in a real life situation at a company;
- Testing different algorithms such as combinations of NB and SVM, as presented by Wang and Manning [13], to possibly improve the classifier's performance.

# 7. CONCLUSIONS

We successfully built a corpus of email messages in Brazilian Portuguese. That was accomplished in association with *Fundação Estudar* a non-profit organization in education that provided us with their email logs.

Based on the corpus created, we produced a study of email classification. We implemented a Naive Bayes and a Support Vector Machine email classifiers and tested precision, recall and F1 statistics for the use of a part-of-speech filter and for the use of a lemmatizer, reaching levels consistent with literature of 87.3% for the F1 score.

The evaluation of performance of the classifier showed that, for email classification in our context, considering only verbs, nouns and adjectives significantly increases performance while adverbs, pronouns, articles, prepositions, conjunctions and interjections tend to influence the classifier negatively. Moreover, it suggested that lemmatizing the corpus reduces classification performance.